\documentclass[12pt]{iopart}

\usepackage{graphicx}  
\begin{document}

\title{Ground state properties of 3d metals from self-consistent GW approach}

\author{Andrey L. Kutepov}

\address{Brookhaven National Laboratory, Upton, NY 11973-5000, USA}

\begin{abstract}
Self consistent GW approach (scGW) has been applied to calculate the ground state properties (equilibrium Wigner-Seitz radius $S_{WZ}$ and bulk modulus $B$) of 3d transition metals Sc, Ti, V, Fe, Co, Ni, and Cu. The approach systematically underestimates $S_{WZ}$ with average relative deviation from the experimental data about 1\% and it overestimates the calculated bulk modulus with relative error about 25\%. It is shown that scGW is superior in accuracy as compared to the local density approximation (LDA) but it is less accurate than the generalized gradient approach (GGA) for the materials studied. If compared to the random phase approximation (RPA), scGW is slightly less accurate, but its error for the 3d metals looks more systematic. The systematic nature of the deviation from the experimental data suggests that the next order of the perturbation theory should allow one to reduce the error.
\end{abstract}

%\noindent{\it Keywords}: GW, FLAPW+LO, equilibrium volume, bulk modulus, transition metals
%\maketitle

\section{Introduction}

The ability of ab-initio theory to predict the ground state properties (GSP) of materials, such as equilibrium volume and bulk modulus, is universally acknowledged as an important test of reliability of the approximations used. In this respect, density functional theory (DFT)\cite{pr_136_B864,pr_140_1133} in its local density approximation (LDA)\cite{prl_45_566,cjp_58_1200} or in the generalized gradient approximation (GGA)\cite{prb_33_8800,prl_77_3865} has a long and successful history. However, DFT has been less successful in describing the excited properties of solids. Other approaches, based on many body perturbation theory (MBPT), such as Hedin's GW method \cite{pr_139_A796}, have been used to study the excitations. Obviously, even from a purely aesthetic point of view, the situation, when one has to use one theoretical framework for GSP and another framework for excited properties, cannot be acknowledged as satisfactory. In this respect, considerable progress has been made in extending the ideas of the DFT to the time-dependent phenomena which resulted in construction of the time-dependent density functional theory (TDDFT) \cite{prl_52_997,pra_35_442,pra_31_1950}. However, the problem of finding the good approximations for the exchange-correlation kernel of the TDDFT is even more difficult than the search for the good approximations for the exchange-correlation functional in standard DFT.

Publications extending the many-body based methods for calculating the GSP for materials are not numerous. Whereas it is rather straightforward theoretically to study the GSP from the many-body perspective, the progress in the field was witnessed mostly for atomic and molecular systems \cite{jcp_130_114105,prb_88_075105}. Only recently, there was progress in extending the many-body methodologies for studying the GSP of solids \cite{prb_66_245103,prl_103_056401,prl_105_196401,prb_87_214102}. Namely, the random phase approximation (RPA) for the correlation energy has been used in the Luttinger and Ward functional \cite{pr_118_1417} for the total energy. The RPA calculations were performed non-self-consistently with one-electron energies and orbitals from GGA calculation as an input for the functional.

This work focuses on the application of the fully self-consistent many-body GW approach (scGW) for studying the GSP of solids.
A few years ago we successfully applied scGW to calculate GSP of simple materials (Na, Al, and Si) \cite{prb_80_041103}. In this work, the scGW approach is used to calculate the equilibrium volume (the corresponding Wigner-Seitz radius is actually presented in the table and in the figures below) and bulk modulus of the transition 3d metals.
Despite their importance in technological applications, systematic theoretical studies of GSP of 3d transition metals are not numerous. Kokko and Das \cite{jcm_10_1285} have compared the performance of LDA and GGA using the linear muffin tin orbitals method (LMTO) in the atomic sphere approximation (ASA). Similar comparison was done later by Haas et al. \cite{prb_79_085104} using full-potential linear augmented plane wave method with local orbitals (FLAPW+LO).
Janthon et al. \cite{jctc_10_3832} studied different flavors of GGA and hybrid functionals. Recently, Schimka et al. \cite{prb_87_214102} studied the GSP of 3d metals using the projector augmented wave (PAW) method. They compared different approximate DFT functionals and RPA implemented in one-shot style with an input from GGA calculations.

Another objective of this work is to emphasize the difference between DFT and scGW in the rate of convergence of the calculated GSP with respect to the size of the basis set. Particularly, in FLAPW+LO methodology (which is used in this work), it is important to monitor the convergence with respect to the number of local orbitals (LO) included in the basis set. The importance of this issue for the GW-based approaches was reported earlier in the context of the calculated band gaps \cite{prb_74_045104,prb_83_081101,prb_84_039906,prb_93_115203,prb_94_035118} and in the context of optimized-effective-potential and response functions \cite{prb_83_045105,prb_85_245124,prb_92_245101}.

\section{Method}
\label{GW}

Let us first outline the basic formulas of  the scGW method described
earlier in Refs. \cite{prb_85_155129,arx_1606_08427}.  In a self-consistent calculation, one has to perform a certain number of iterations until the self-consistency is achieved. In the scGW method, the input for every
iteration is the Green function $G$. For a given G, we perform a few intermediate steps, such as the calculation of polarizability
\begin{equation}  \label{GW_1}
P(12;\tau)=-G(12;\tau)G(21;\beta-\tau),
\end{equation}
the screened interaction
\begin{equation}  \label{GW_2}
W(12;\nu)=V(12)+\int d(34) V(13)P(34;\nu)W(42;\nu),
\end{equation}
the self-energy 
\begin{equation}  \label{GW_3}
\Sigma(12;\tau)=-G(12;\tau)W(21;\tau),
\end{equation}
and new Green's function
\begin{equation}  \label{GW_4}
G(12;\omega)=G_{0}(12;\omega)+\int d(34) G_{0}(13;\omega)\Sigma(34;\omega)G(42;\omega).
\end{equation}

In the above equations, the numbers (e.g. 12, 34) are used to denote space variables, $\tau$ is the Matsubara's time, $\nu$ and $\omega$ are the Matsubara's frequencies (bosonic and fermionic correspondingly), $G_{0}$ is the Green's function in Hartree approximation, and $\beta$ is the inverse temperature. Spin index is not shown in the above equations for simplicity.

In order to evaluate free energy needed to study the GSP, the finite temperature variational expression for the grand potential \cite{NMBTQS_MI} based on the work by Luttinger and Ward \cite{pr_118_1417} is evaluated first:

\begin{equation}\label{om1}
\Omega=\Omega_{x}+\Phi_{c}-Tr[G^{-1}_{x}G-1]+Tr[\ln G-\ln G_{x}],
\end{equation}
where the exchange part (first term on the right hand side) and the correlation part (the remaining terms on the rhs) were separated for convenience. The exchange part of the grand potential is:

\begin{equation}\label{om2}
\Omega_{x}=-\frac{1}{\beta N_{\mathbf{k}}}\sum_{\alpha \mathbf{k}\lambda} \ln 
[1+e^{-(\varepsilon^{\alpha\mathbf{k}}_{\lambda}-\mu)\beta}]+E_{coul}-Tr (V^{H}G)-\frac{1}{2} Tr (\Sigma^{x}G),
\end{equation}
where $\alpha$ is the spin, $\mathbf{k}$ enumerates k-points in the Brillouin zone, $\lambda$ is the band index, $\beta$ is the inverse temperature, $N_{\mathbf{k}}$ is the total number of k-points in the Brillouin zone, $\mu$ is the chemical potential. $E_{coul}$ represents the energy of the Coulomb interaction (sum of nuclear-nuclear, electron-nuclear, and classical electron-electron). $V^{H}$ is the electronic Hartree potential, $\Sigma^{x}$ is the exchange (frequency-independent) part of self energy, $G_{x}$ is the exchange part of Green's function, and $G$ is the total Green function.

For the remaining (correlation) terms in Eq.(\ref{om1}), the following expressions were used in practical calculations:

\begin{equation}\label{om3}
\Phi_{c}=\frac{1}{4}Tr \{ \ln(1-PV)(1-VP) +PV+VP\},
\end{equation}

\begin{equation}\label{om4}
Tr[G^{-1}_{x}G-1]=Tr[\Sigma_{c}G],
\end{equation}

\begin{equation}\label{om5}
Tr[\ln G-\ln G_{x}]=Tr_{0} \sum_{\omega>0}[\ln [G(\omega)G^{+}(\omega)-G_{x}(\omega)G^{+}_{x}(\omega)].
\end{equation}

In the equations above, $P$ is the irreducible polarizability, $V$ is the bare Coulomb interaction, $\Sigma_{c}$ is the correlation self energy. Matrices in Eqs. (\ref{om3}) and (\ref{om5}) have been transformed to the hermitian form in order to facilitate the evaluation of the logarithm. Taking the trace ($Tr$) means the summation over the k-points in the Brillouin zone, spins, Matsubara's frequencies, and diagonal matrix elements. Symbol $Tr_{0}$ means that frequency summation is excluded from the evaluation of trace. Having evaluated the grand potential, the free energy simply is $F=\Omega-\mu N$, with $N$ being the number of electrons. In order to obtain the equilibrium volume and the corresponding bulk modulus, free energy was evaluated on the equidistant mesh of five volumes arranged near the theoretical minimum of free energy. The obtained function $F(V)$ was approximated by quadratic polynomial of $V$ which allowed one to get the position of the minimum (equilibrium volume) and the second derivative (related to the bulk modulus).

Bloch band states of the effective exchange problem (corresponds to the approximation $\Sigma^{c}=0$) were used to represent $G$, $V^{H}$, $\Sigma^{x}$, and $\Sigma^{c}$. The polarizability and the Coulomb interaction were expressed in the mixed product basis \cite{prb_76_165106,arx_1606_08427}. FLAPW+LO method was used to solve the one-electron problems (Kohn-Sham equation in LDA and effective exchange hamiltonian \cite{prb_85_155129}). Because of the importance of the local orbitals for the present study and because of slightly different definitions of them in the literature \cite{prb_93_115203,prb_64_195134,prb_43_6388,Singh94}, a few specifics of their implementation in the present study are given here. In this work, one local orbital was defined for every principal quantum number excluding the principal quantum number for which the LAPW augmentation was performed. It was constructed as a linear combination of a solution of radial equation and its first- and second energy derivatives with the requirement for the local orbital to be normalized and to have vanishing value and slope at the boundary of the corresponding muffin-tin (MT) sphere. Radial equations were solved with an energy parameter equal to the band center (semi-core states) and with energy parameter corresponding to the logarithmic derivative D equal to \textit{-l-1} (\textit{l} is orbital momentum quantum number) for valence and excited states. The number of LO's for every \textit{l} was defined by specifying (from the convergence requirement) the maximal principal quantum number $n_{max}$. For the materials studied in this work $n_{max}$ was 10-12. The maximal value of the orbital momentum in MT spheres was equal 8. The number of the augmented plane waves was 70-100/atom for the studied materials. Integrations over the Brillouin zone were performed using the mesh $8\times 8\times 8$ for the materials with the body-centered cubic structure (V and Fe) and with the face-centered cubic structure (Ni and Cu). The mesh was $6\times 6\times 4$ for the materials with the hexagonal close-packed structure (Sc, Ti, Co). The size of the mixed product basis is related to the quality of LAPW+LO basis and was varied in the range 560-1500 functions per atom. All calculations were performed for the electronic temperature 1000K. Metals Fe, Co, and Ni were assumed to be ferromagnetic. The calculations for the rest of the materials have been performed without spin polarization. Two 3d metals, Mn and Cr, were excluded from this study. Mn has too many atoms (58) per unit cell in its ground state which at the moment is beyond of the capability of scGW approach. Cr has complicated magnetic structure effectively making the calculations very time consuming.

\section{Results and discussion}
\label{res}

\begin{figure}[t]
\centering
\includegraphics[width=9.0 cm]{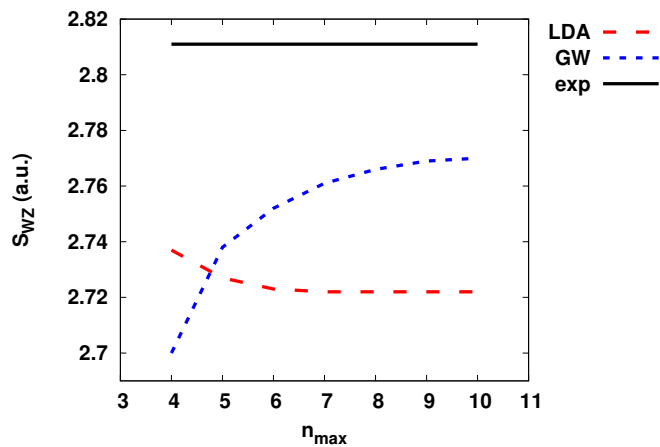}
\caption{Convergence of the calculated equilibrium Wigner-Seitz radius ($S_{WZ}$) of vanadium with respect to the maximal principal quantum number $n_{max}$. The experimental Wigner-Seitz radius corrected for zero-point vibrational effects \cite{prb_87_214102} is shown as the horizontal line.} \label{swz_n}
\end{figure}

\begin{figure}[b]
\centering
\includegraphics[width=9.0 cm]{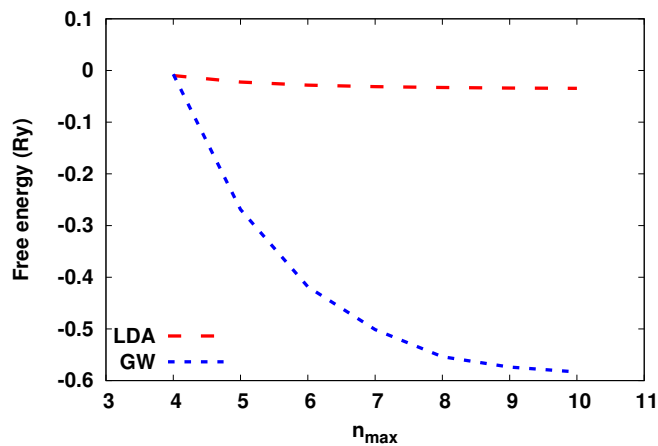}
\caption{Convergence of the calculated free energy of vanadium with respect to the maximal principal quantum number $n_{max}$. For plotting purposes, the constant 1897.37Ry has been added to the free energy in scGW approximation, and the constant 1894.78Ry has been added to the LDA result.} \label{e_conv_n}
\end{figure}

\begin{figure}[t]
\centering
\includegraphics[width=9.0 cm]{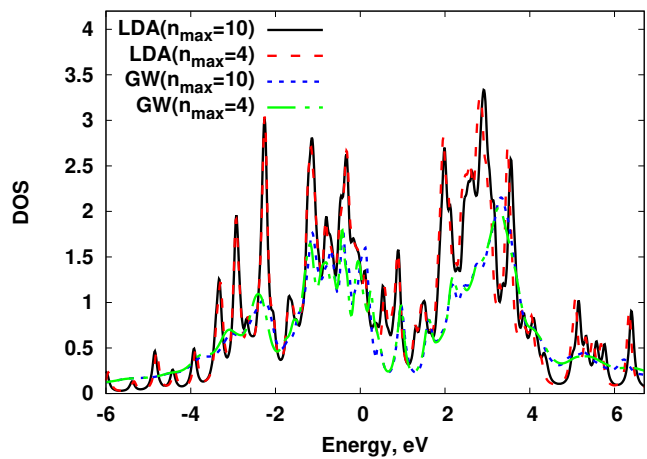}
\caption{Density of states (spectral function) evaluated in LDA (scGW) with the different number of high energy local orbitals included.} \label{dos_conv_n}
\end{figure}

\begin{figure}[b]
\centering
\includegraphics[width=9.0 cm]{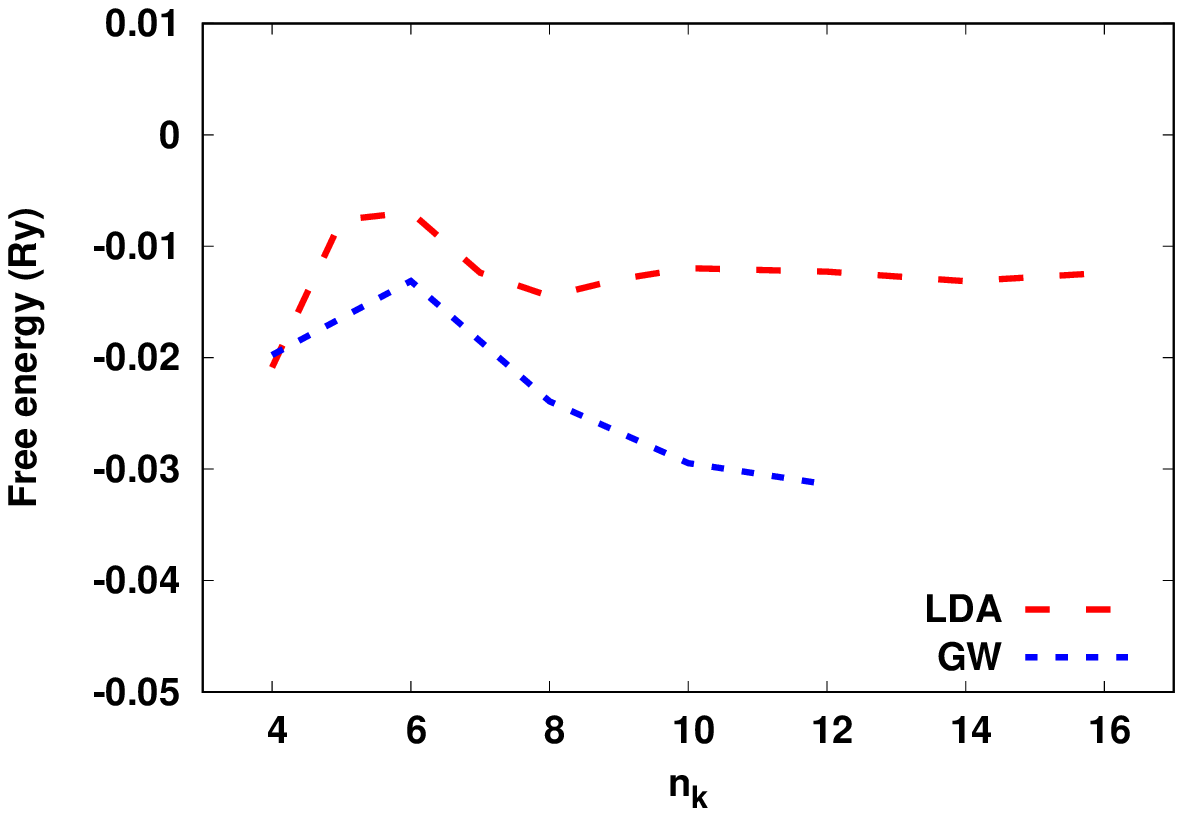}
\caption{Convergence of the calculated free energy of vanadium with respect to the number of k-points in the Brillouin zone. For plotting purposes, the constant 1897.92Ry has been added to the free energy in scGW approximation, and the constant 1894.80Ry has been added to the LDA result.} \label{e_conv_k}
\end{figure}

Let us begin the discussion of results by stressing the essential difference in how one should perform sc LDA or scGW calculations. Total energy in LDA depends only on the quality of the occupied states description. As such, the convergence of the total energy with respect to the size of a basis set in LDA is relatively fast. On the contrary, in the scGW calculations, one has to perform the summation not only over the occupied states but also over an infinite number of unoccupied states. It makes the convergence of the total energy with respect to the size of a basis set very slow in the scGW approximation. This problem was already documented in atomic calculations. For example, Sakuma and Aryasetiawan \cite{pra_85_042509} have shown that in order to get the convergent total energy of the hydrogen atom (the simplest one!) in the GW approximation, one has to sum up not only over bound states but also over an infinite number of the continuum states. In this work, it is shown that similar problem exists for crystalline materials. Namely, working with the LAPW+LO basis set, one can observe very slow convergence of the total energy (and, correspondingly, the calculated GSP) with respect to the number of LO's included in the basis set. Figure \ref{swz_n} demonstrates this fact using vanadium as an example. As one can see from Figure \ref{swz_n}, the equilibrium Wigner-Seitz radius obtained in LDA is sufficiently well converged with just one or two LO's (per angular momentum) added to the LAPW basis. The situation is, however, totally different in the scGW case. One has to include all LO's up to $n_{max}=10$ (up to 7 LO's per angular momentum) in order to stabilize the calculated $S_{WZ}$. It is important to mention that the free energy itself shows even worse convergence in the scGW method than $S_{WZ}$ does (Fig. \ref{e_conv_n}). It is, obviously, in accord with the above mentioned need to include the continuum states in the case of the hydrogen atom \cite{pra_85_042509}. Fortunately, the error related with the neglect of even higher energy excited states in the present calculations becomes almost independent on volume which makes it possible in practice to obtain reasonably converged ground states properties. The GSP of the rest of the materials studied in this work show similar convergence with respect to $n_{max}$ as vanadium does. There is only a slight tendency towards slower convergence when one goes from V to Cu (corresponds to an increase in the number of electrons per atom). For example, in nickel and in copper, the convergence of $S_{WZ}$, similar to the convergence observed in vanadium, is achieved with $n_{max}=12$. The convergence is good enough for studying the general tendencies (as in this work) but not sufficient for studying, for example, the elastic constants, where the small differences in $F$ have to be evaluated accurately. However, if one is interested only in the one-electron spectra near the Fermi level, the quality of the corresponding spectral functions is already good enough even for $n_{max}=4$ as one can judge from Figure \ref{dos_conv_n}. It is obvious from Fig. \ref{dos_conv_n} that the difference in the results obtained with $n_{max}=4$ and $n_{max}=10$ is very similar in LDA and scGW, and it is acceptable in most cases. One can explain this phenomenum by noticing that the free energy in the GW approximation depends not only on the low energy part of the self energy (as the one-electron spectrum does) but also on the high-energy part of the self energy and, thus, effectively includes the sum over exited states twice (every element of the self energy is already a sum over intermediate states, including the high-energy states, plus, we have to sum up an infinite number of elements in equations (\ref{om4},\ref{om5})).

\begin{table}[b]
\caption{Theoretical and experimental equilibrium Wigner-Zeitz radius ($S_{WZ}$) and bulk moduli ($B_{0}$). The experimental data corrected for zero-point vibrational effects have been cited from Refs. \cite{prb_87_214102,jctc_10_3832}.} \label{v0b0}
\begin{center}
\begin{tabular}{@{}c c c c c c c} &\multicolumn{3}{c}{$S_{WZ}$ (a.u.)} &
\multicolumn{3}{c}{$B_{0}$ (MBar)}\\
 &LDA &scGW & Experiment &LDA  &scGW   &  Experiment \\
\hline
Sc &3.32 &3.39 & 3.42 & 0.73 & 0.74 & 0.56\\
Ti &2.97 &3.01  &3.05  &1.29  &1.45  &1.08 \\
V &2.73 &2.77 &2.81  &2.47  &2.15  &1.59\\
Fe &2.56 &2.64  &2.65  &2.84  &1.90   &1.70 \\
Co &2.53 &2.57  &2.60  &2.78  &2.33  &1.93\\
Ni &2.52 &2.55  &2.59  &2.84  &2.46  &1.86\\
Cu &2.61 &2.62  &2.66  &1.78  &1.90     &1.40
\end{tabular}
\end{center}
\end{table}

\begin{figure}[t]
\centering
\includegraphics[width=9.0 cm]{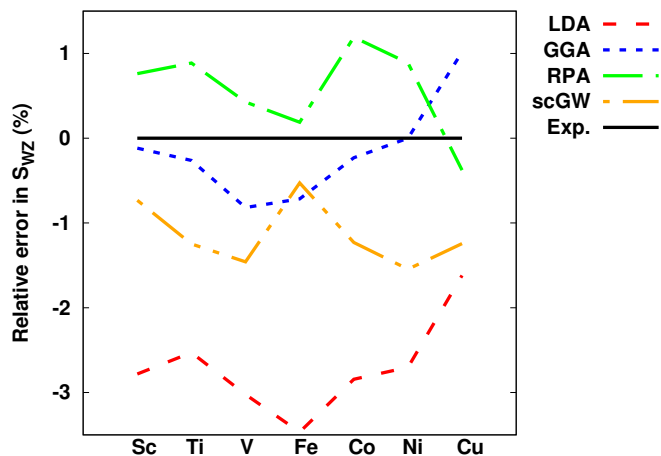}
\caption{Relative error in the calculated equilibrium Wigner-Seitz radius ($S_{WZ}$) of the 3d transition metals. LDA and scGW results are from this work. GGA and RPA results are from the Ref.\cite{prb_87_214102}. The experimental data corrected for zero-point vibrational effects have been cited from Ref. \cite{prb_87_214102}.} \label{swz}
\end{figure}

\begin{figure}[b]
\centering
\includegraphics[width=9.0 cm]{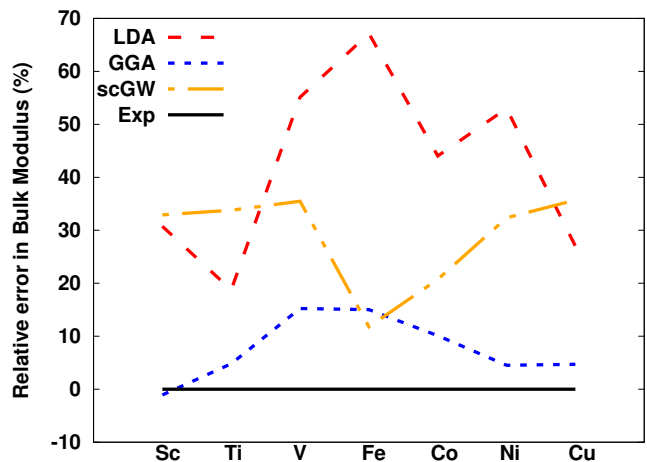}
\caption{Relative error in the calculated bulk moduli of the 3d transition metals. LDA and scGW results are from this work. GGA results and the experimental data corrected for zero-point vibrational effects have been cited from the Ref. \cite{jctc_10_3832}.} \label{bm}
\end{figure}

\begin{figure}[t]
\centering
\includegraphics[width=9.0 cm]{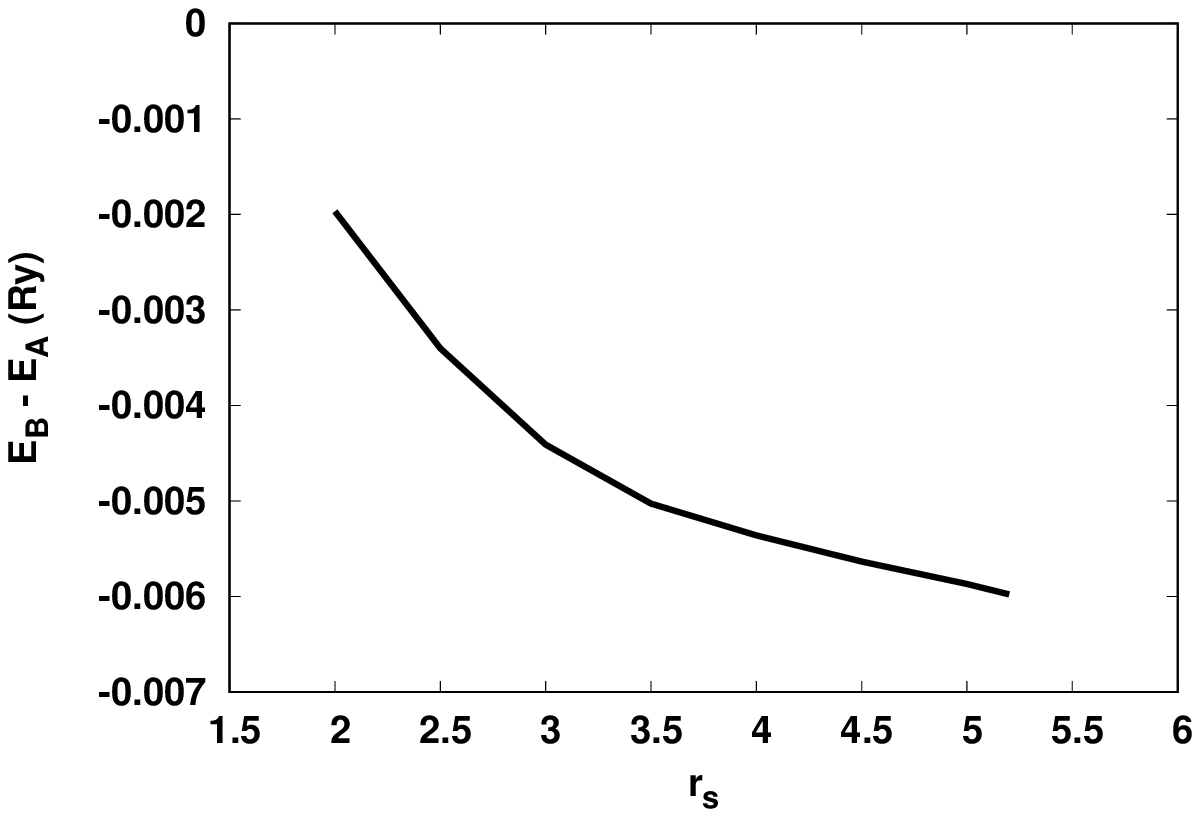}
\caption{Difference in the total energy of the electron gas evaluated in the scGW+Vertex approximation (scheme B from Ref. \cite{prb_94_155101}) and in scGW (scheme A) approximation as a function of $r_{s}$ (radius of the sphere containing one electron). The data have been taken from our work on the electron gas \cite{prb_96_035108}.} \label{e_vrt}
\end{figure}

The convergence with respect to the number of k-points in the Brillouin zone doesn't show dramatic difference between LDA and scGW approaches (Figure \ref{e_conv_k}). Taking into account the difference in scales in the Figs. \ref{e_conv_n} and \ref{e_conv_k}, one can estimate that the uncertainty in the calculated free energy related to the finite number of k-points is a few times smaller than the uncertainty related to the finite size of the basis set.

Table \ref{v0b0} presents the calculated $S_{WZ}$ and the corresponding bulk modulus $B$ (evaluated at the theoretical equilibrium volume) for the materials studied in this work. Figures \ref{swz} and \ref{bm} show the relative deviation of the calculated $S_{WZ}$ and $B$ from the experimental data corrected for the zero-point lattice vibrations. The results obtained in GGA and RPA (both are from the Ref. \cite{prb_87_214102}) are given for the comparison as well (RPA results are available only for $S_{WZ}$). As one can see from Figures \ref{swz} and \ref{bm}, the improvements in the results obtained in the scGW (as compared to the LDA results) are noticeable. The deviation of scGW results from the experimental data are systematic ($S_{WZ}$ is always underestimated with an average relative error of approximately 1\%, whereas $B$ is overestimated by about 25\%). However, GGA shows better performance than scGW (though might be less systematic). RPA overestimates $S_{WZ}$ with an average relative error slightly less than the relative error obtained in the scGW approximation. It is interesting that two many-body based methods (RPA and scGW) result in the smallest relative error in $S_{WZ}$ when applied to the iron. This fact suggests that Fe is the less correlated material among the metals studied in this work. As a certain support of this statement, one can mention the recent work by Sponza \textit{et al} \cite{prb_95_041112} where it was shown that the one-electron spectrum of Fe can be well described in sc quasiparticle GW approximation (QSGW) \cite{prb_76_165106}, whereas the strong spin fluctuations in Ni required one to use additional modelistic approaches to reproduce the experimental ARPES data. 

The advantage of the scGW approach as compared to the LDA or the GGA is that there is well defined way to improve the accuracy. Namely, one can apply self-consistent vertex corrected schemes which were implemented and successfully applied for the spectral properties of crystalline materials in Refs. \cite{prb_94_155101} and \cite{prb_95_195120} recently. They are, however, quite expensive computationally, and the studying of their effect on the GSP of solids can be a subject of a separate research. In order to support the idea that vertex corrections should increase the calculated equilibrium volume, the difference in the total energy of the electron gas between the results obtained in the vertex corrected GW calculations and in the GW calculations without vertex corrections is presented in Figure \ref{e_vrt}. As one can see, in the range of metallic densities ($r_s$ between 2 and 5), vertex corrections make the total energy more negative when the volume increases.

\begin{table}[t]
\caption{Calculated (in LDA and in scGW) magnetization of the ferromagnetic 3d metals compared to the experimental data. The theoretical results are given for the calculated equilibrium volumes. The experimental data are cited from Ref. \cite{prl_96_226402}.} \label{mmom}
\begin{center}
\begin{tabular}{@{}c c c c} &LDA &scGW & Experiment\\
\hline
Fe &2.00 &2.93  &2.22\\
Co &1.52 &1.93  &1.60\\
Ni &0.57 &0.67  &0.60
\end{tabular}
\end{center}
\end{table}

Whereas the studying of the magnetic properties was not the main goal of this work, it is interesting from the methodological point of view to see what the scGW approach gives for the most basic of them. Table \ref{mmom} presents spin magnetic moments of Fe, Co, and Ni, evaluated with the scGW method and compares them with the LDA results and with the experimental data. The moments obtained in the LDA are very close to the experimental data with only a slight underestimation of them (but notice that the moments were evaluated at the theoretical equilibrium volume). The scGW approach overestimates the calculated moments considerably. This fact suggests that one has to go diagrammatically beyond the scGW approximation in order to improve the many-body prediction of the magnetic properties.

In conclusion, self-consistent GW approximation was applied to study the ground state properties of 3d transition metals. It was demonstrated that the approach is more accurate than the LDA, but less accurate when compared to the GGA. The results suggest that one has to supplement the GW approximation with the vertex corrections in order to improve the accuracy of the many-body theory when it is applied to the calculation of the GSP. The study also revealed the importance of the proper convergence with respect to the size of the basis set in scGW calculations of the GSP, which was reported earlier in the context of the one-electron spectra and susceptibilities.

\section*{Acknowledgments}
\label{acknow}

This work was   supported by the U.S. Department of energy, Office of Science, Basic
Energy Sciences as a part of the Computational Materials Science Program.

\section*{References}
\label{ref}

\bibliographystyle{iopart-num}
%\bibliography{Method,Actinides}

\providecommand{\newblock}{}

\end{document}